\documentclass{appolb}
\usepackage{subfigure}
 \usepackage[dvips]{graphicx}
 \DeclareGraphicsExtensions{.eps, .jpg}
\usepackage{amsmath,amssymb,amsbsy}
\def\vec#1{\boldsymbol{#1}}
\def\nuc#1#2{{\ensuremath{{}^{#2}\mathrm{#1}}}}
\def\clu#1#2#3{{\ensuremath{{}^#2\mathrm{#1}_{#3}}}}
\def\be{\begin{equation}}
\def\ee{\end{equation}}
\def\bel#1{\begin{equation}\label{#1}}
\begin{document}
\author{Jean-Marc Richard
\thanks{\texttt{email: jean-marc.richard@lpsc.in2p3.fr}}
\thanks{Talk given at the Workshop ``Excited QCD'',  February 8--14,  2009,  Zakopane, Poland} 
\address{ Laboratoire de Physique Subatomique et Cosmologie,\\[2.5pt]
Universit\'e Joseph Fourier, CNRS-IN2P3, INPG,\\
53, avenue des Martyrs, Grenoble, France}}
%
\title{\bf Steiner-tree confinement and tetraquarks}
\maketitle
\begin{abstract}
The linear confinement in quarkonium is generalised as a minimal tree, with interesting geometrical properties.
This model binds tetraquarks more easily than the additive model used in earlier investigations. 
\end{abstract}
\section{Introduction}\label{se:intro}
The stability of compound systems is a delicate issue.
In atomic physics, binding $\mathrm{H}^-$ cannot be established using any Hartree type of wave function $f(r_1)f(r_2)$. The positronium molecule, $\mathrm{Ps}_2=(e^+,e^+,e^-,e^-)$ is rather weakly bound below the two-atom threshold. The configuration $(p,e^+,e^-)$ is unstable, but it is stabilised by  a second electron, leading to the positronium hydride, below the $\mathrm{H+Ps}$ threshold. There are many other examples.

In molecular physics, the dimer \clu{He}{4}{2} is barely bound, while the 
 higher \clu{He}{4}{n}  clusters with $n\ge 3$ have much deeper binding per atom, illustrating the Thomas effect \cite{PhysRev.47.903}. In the same He--He potential,  the lighter  \clu{He}{3}{2} dimer is unbound, even for spin singlet. It 
was attempted to calculate  \clu{He}{3}{n} with $n = 3,\ldots10$, which were 
found unbound. Every reasonable physicist would have given up, but fortunately%
\footnote{%
\emph{Il faut avoir un peu de folie, qui ne veut avoir plus de sottise.} (One should be a little foolish to avoid craziness.) Michel de Montaigne}
some colleagues were curious enough to push the calculation to higher $n$, and discovered that
\clu{He}{3}{n} is bound for $n\gtrsim 35$, or less if there are also some \nuc{He}{4} admixed \cite{Barranco}.

In nuclear physics, some improbable configurations  turn out eventually stable, due to the pairing interaction. For instance, while \nuc{He}{5} is unstable. \nuc{He}{6} is stable. Seen as a $(\alpha,n,n)$ three-body system,  it is \emph{Borromean}, as none of the two-body subsystem is bound. Even more remarkable is that, though \nuc{He}{7} is unstable, one finds again stability for \nuc{He}{8},  something like a small drop of neutron star around two protons. 

In hadron physics, too,  exotics probably require a sophisticated arrangement of the quarks and antiquarks. After the wave of baryonium candidates, which were not confirmed, and the absence of firmly-established light pentaquarks, some scepticism now prevails about multiquark spectroscopy. However, many sectors remain to be explored, for instance, mesons with charm 2 or with beauty $-2$.
%

The best known category of multiquarks includes the deuterium and other nuclei, but, due to the repulsive core of nuclear forces, each nucleon keeps it identity. The Yukawa  mechanism is by no means restricted to the nucleons, and acts between several other hadron pairs, provided they contain light quarks.
In particular, the pion-exchange is attractive in the $D\overline{D}{}^*+D^*\overline{D}$ channel. As compared to the deuteron, the potential is weaker but experienced by heavier constituents. Hence this approach was considered as successful when the $X(3872)$ was discovered just above the $D\overline{D}{}^*$ threshold. However, there is no hard core here, and one cannot escape the short-range forces, which presumably result from a direct interaction between the heavy and light quarks.
For a recent review, and refs.~to the pioneering papers, see, e.g., \cite{Swanson:2006st}.

Some years ago, an interesting coherence has been discovered in the colour--spin operator entering the spin-spin interaction known as chromomagnetism \cite{Jaffe:1999ze}. Some expectations values are larger (and with the right sign for inducing attraction) for some multiquark configurations than for the sum of the decay products. This is the case in particular for the H$(uuddss)$ as compared to $\Lambda+\Lambda$, or for the 1987-vintage pentaquark \cite{Gignoux:1987cn}, $(\overline{Q}qqqq)$, as compared to $(Q\bar{q})+(qqqq)$, where $(qqqq)$ denotes $(uuds)$ or permutations. The first estimates were carried out with unbroken flavour SU(3)$_\text{f}$ symmetry, and assuming that the quark--quark correlation is the same for multiquarks as for ordinary hadrons. Unfortunately, breaking the flavour symmetry benefits more to the threshold than to the H or P and thus spoils bindings. Also multiquarks are more dilute systems than ordinary hadrons. Thus the strength of spin--spin forces is appreciably reduced.  Altogether, the H and the P seem likely unbound when all effects are taken into account.

If the chromomagnetic interaction, once rather promising, turns out disappointing, it is natural to address the possibility of binding using the \emph{chromoelectric interaction}, in particular through its properties under symmetry breaking and its $N$-body character (this is not a sum of pairwise terms).
\section{Flavour independence and symmetry breaking}\label{se:sym-break}
Symmetry breaking is known to lower the ground state of any Hamiltonian. The simplest example is $p^2+x^2+\lambda x$ in one dimension, with the lowest energy $E(\lambda)=1-\lambda^2/4\le E(0)$. More generally if
\bel{eq:sym-b1}
H=H_0+\lambda\,H_1~,
\ee
where $H_0$ is even under some symmetry, and $H_1$  odd, then the variational principle with the even ground state solution of the $\lambda=0$ case as a trial wave function gives 
\bel{eq:sum-b2}
E(\lambda)\le E(0)~.
\ee
However, in most cases, the threshold energy  benefits more  from this effect, and symmetry breaking deteriorates stability. This is what happens with the H vs. $\Lambda+\Lambda$ in the chromomagnetic model, when SU(3)$_\text{f}$ is broken. Another example is the breaking of permutation symmetry in molecules. The equal mass case, say $(\mu^+,\mu^+,\mu^-,\mu^-)$, a rescaled version of the positronium molecule, is bound by about 3\% below the threshold for dissociation into two neutral atoms (internal annihilation is neglected here). Consider now the asymmetric configuration $(M^+,m^+,M^-,m^-)$, with $\mu^{-1}$ average of $M^{-1}$ and $m^{-1}$. The Hamiltonian can be written as
\bel{eq:sym-b3}
H(M^+,m^+,M^-,m^-)=H(\mu^+,\mu^+,\mu^-,\mu^-)+\left[\frac{1}{4M}-\frac{1}{4m}\right]
(\vec p_1^2-\vec p_2^2+\vec p_3^2-\vec p_4^2)~,
\ee
demonstrating that $E(M^+m^+,M^-,m^-)\le E(\mu^+,\mu^+,\mu^-,\mu^-)$. However, meanwhile, the threshold evolves from $2\,E_2(\mu,\mu)$ to $E_2(M,M)+E_2(m,m)$ which is lower. Indeed, it has been shown from accurate four-body calculation that stability holds only for 
$1/2.2\lesssim M/m\lesssim 2.2$. See, e.g., \cite{Armour}.

However, a miracle occurs when breaking charge conjugation. Consider a mass configuration $(M,M,m,m)$ and a potential that does not depend on the masses, as the Coulomb potential in atomic physics, or the static potential in QCD (flavour independence). Then 
\bel{eq:sym-b4}
H(M,M,m,m)=H(\mu,\mu,\mu,\mu)+\left[\frac{1}{4M}-\frac{1}{4m}\right]
(\vec p_1^2+\vec p_2^2-\vec p_3^2-\vec p_4^2)~,
\ee
Again, the ground-state energy is lowered by the odd term, but the threshold energy remains \emph{constant}, $2\,E(M,m)=2\, E_2(\mu,\mu)$, as  the two-body energy depends only on the reduced mass.  Thus stability is improved. For molecules, it is, indeed, observed that the relative excess of energy as compared to the threshold, evolves from about 3\% for $M=m$ to nearly $17\%$ for $M\gg m$, e.g., for the hydrogen molecule. 

Similarly, explicit quark model calculations with a flavour-independent potential indicate that stability is reached for $(Q,Q,\bar{q},\bar{q})$ against dissociation into two flavoured mesons if the mass ratio is large enough.

Another miracle is that while the spectroscopy of exotic hadrons is usually a matter for hot controversy, the stability of  $(Q,Q,\bar{q},\bar{q})$ in the large $M/m$ limit  has reached an overall consensus. See, e.g., \cite{Janc:2004qn,Ay:2009zp} for refs. This is also confirmed in a QCD sum-rule calculation \cite{Navarra:2007yw}.

The question now is whether the double charm is heavy enough to get stability. On the experimental side, there are candidates for the double charm baryons from the SELEX experiment, but they are not confirmed in other experiments, yet, and double-charm mesons have never been searched for.  See, e.g.,\cite{Aubert:2006qw} and refs.\  there. On the phenomenological side, explicit four-body calculations have been carried out of $(Q,Q,\bar{q},\bar{q})$ and similar configurations, using potential models tuned to reproduce ordinary mesons and baryons. The usual conclusion is that double charm is not sufficient to bind, and that double beauty would be safer. However, Rosina et al. \cite{Janc:2004qn}, pushing very far a variational calculation with a realistic potential,  got the state $(c,c,\bar{\mathstrut c},\bar{d})$ weakly bound.
\section{Steiner tree of confinement}\label{se:Steiner}
The problem, however, is whether the interquark potential can be safely extrapolated from mesons and baryons to multiquarks. The usual prescription is the colour-additive rule
\bel{eq:add}
V(\vec r_1,\vec r_2,\ldots)=-\frac{3}{16}\sum_{i<j} \tilde\lambda_i.\tilde\lambda_j\,v(r_{ij})~,
\ee
where $v(r)$ is the quarkonium potential. This is perhaps justified for short-range contributions corresponding to colour-octet exchange. But there is not a serious reason, that it should be adequate for confinement. In fact, many years ago, the generalisation of $v(r)=r$ (in units where the string constant is unity) has been proposed by several authors. For refs., see, e.g., \cite{Ay:2009zp}. It has been recently supported by detailed simulations using the lattice QCD \cite{Suganuma:2008ej}. 

\begin{figure}[hbtc]
\vspace{-.2cm}
\centering
\includegraphics[width=.3\textwidth]{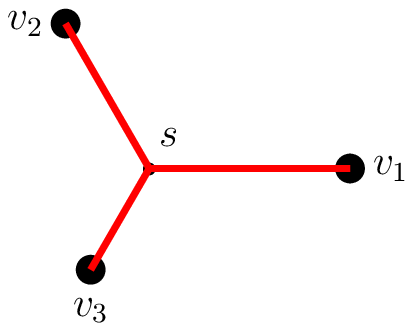}
%
\includegraphics[width=.3\textwidth]{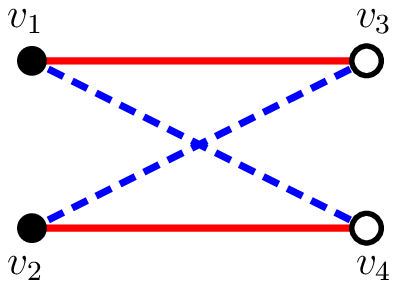}
%
%
\includegraphics[width=.3\textwidth]{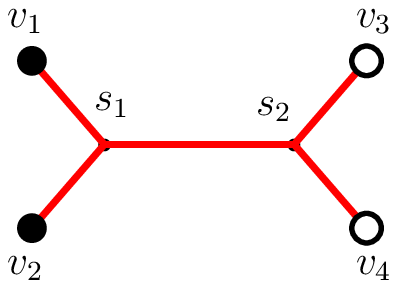}
\vspace{-.2cm}
\caption{\label{fig:bar-tetra} Generalisation of the linear quark--antiquark potential of mesons to baryons (left) and to tetraquarks, where the minimum is taken of the flip--flop (centre) and Steiner tree (right) configurations.}
\end{figure}

For baryons, the extension is the so-called $Y$-shape potential
\begin{equation}\label{Y-shape}
V_3(v_1,v_2,v_3)=\min_s(d_1+d_2+d_3)~,
\end{equation}
where $d_i$ is the distance of the $i^\text{th}$ quark located at $v_i$ ($i=1,\,2,\,3$)  to a junction $s$ whose location is adjusted to minimise $V_3$. This corresponds to the well-known problem of Fermat and Torricelli to link three  points with a minimal network.  See Fig.~\ref{fig:bar-tetra}.

For tetraquarks, the potential, also pictured in  Fig.~\ref{fig:bar-tetra}, reads (with $d_{ij}=\| v_i v_j\|$)
\begin{equation}\label{tetra-pot}\begin{aligned}
U&=\min \left\{ d_{13}+d_{24},d_{14}+d_{23},V_4 \right\}~,\\
\qquad
V_4&=\min_{s_1,s_2}\left(\, \|v_1s_1\| + \| v_2s_1\|+ \| s_1s_2\|+\| s_2v_3\|+\| s_2v_4\|\,\right )~,
\end{aligned}\end{equation}
with the minimum of the two possible quark--antiquark links,  sometimes referred to as the ``flip--flop'' model, and the connected flux tube, itself minimised by varying the location of the Steiner points $s_1$ and $s_2$.

A first investigation using the potential (\ref{tetra-pot}) concluded to the absence of exotics \cite{Carlson:1991zt}.
 However,  Vijande et al.~\cite{Vijande:2007ix} used a more systematic variational expansion of the wave function and in their numerical solution of the four-body problem, found a stable tetraquark ground state. Moreover, unlike \cite{Carlson:1991zt}, they considered the possibility of unequal masses, and found that stability  improves if the quarks are heavier (or lighter) than the antiquarks, in agreement with the earlier argument about favourable symmetry breaking.
  
Recently, an inequality has been derived for this tetraquark potential, 
 \bel{eq:boundU2}
U\le \frac{\sqrt3}{2}\left(\| x\| +\| y\| \right) + \| z \| ~.
\ee
The four-body Hamiltonian with this upper bound splits into three simple one-variable Hamiltonians with linear confinement, and can be solved analytically. In the limit of large $M/m$, the stability below the dissociation threshold is recovered \cite{Ay:2009zp}.
\section{Outlook}
The recent progress in understanding confinement has inspired more realistic quark potentials which have been used for studying multiquark spectroscopy. Remarkably, this new modelling of confinement gives better attraction than the conventional colour-additive models, and thus predicts a richer spectrum of exotics.
This is  confirmed by a detailed study of $(Q,Q,\bar{q},\bar{q})$ configurations. It is our intend to apply the same dynamics to pentaquark and dibaryon configurations.

\subsection*{Acknowledgements} I would like to thank the organisers for the pleasant and stimulating atmosphere of the Workshop, and M.~Asghar for useful comments.


\end{document}